\documentclass{aa}
\usepackage{graphicx}
\usepackage{natbib}
\usepackage[cmex10]{amsmath}
\bibpunct{(}{)}{;}{a}{}{,} 
\begin{document}
\def\reffont{}
   \title{Beam squint and Stokes V with off-axis feeds}

  \author{ Juan M. Uson
  \and W.~D.~Cotton
	}

  \institute{National Radio Astronomy 
	Observatory\thanks{The National Radio Astronomy Observatory
		(NRAO) is operated by Associated Universities Inc.,
		under cooperative agreement with the National Science 
                Foundation.}, 
	520 Edgemont Road, Charlottesville, VA 22903-2475, USA
         }
   \offprints{J.~M.~Uson}

   \date{Received 4 February 2008 / Accepted 9 May 2008}

   \abstract{
Radio telescopes with off-axis feeds, such as the (E)VLA, suffer
from ``beam squint'' in which the two orthogonal circular polarizations sampled
have different pointing centers on the sky.
Its effects are weak near the beam center but become increasingly
important towards the edge of the antenna power pattern where gains in the
two polarizations at a given sky position are significantly different.
This effect has limited VLA measurements of
circular polarization (Stokes~V) and introduced dynamic
range limiting, wide-field artifacts in images made in Stokes~I.
We present an adaptation of the visibility-based deconvolution CLEAN method
that can correct this defect ``on the fly'' while imaging, correcting as well the
associated self-calibration.
We present two examples of this technique using the procedure ``Squint'' within
the Obit package which allows wide-field imaging in Stokes~V and reduced
artifacts in Stokes~I.  We discuss the residual errors in these examples as well as
a scheme for future correction of some of these errors.
This technique can be generalized to implement temporally- and
spatially-variable corrections, such as pointing and cross-polarization leakage errors.
   }

   \maketitle

\section{Introduction}
The enhanced sensitivity of the EVLA will allow it to study sources
much weaker than are currently observed.
For frequencies at the lower end of the EVLA range this will generally
require imaging the full primary beam of the antenna pattern, or at
least those portions containing the stronger sources in the field in
order to remove their sidelobes from the region of interest.
In this regime, numerous effects of minor importance to observations
of strong sources near the antenna pointing position must be
understood and removed in order to allow the EVLA to reach its potential.
``Beam squint'' resulting from the off-axis arrangement of the feeds
is one of these effects.

\begin{figure}
   \centering
   \includegraphics[height=3.5in]{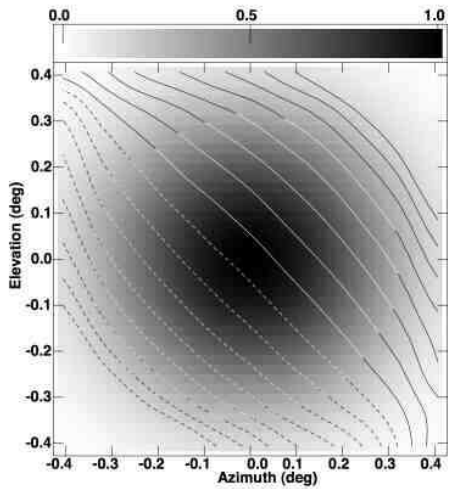}
\caption{ 
The VLA primary antenna pattern as measured during the NVSS survey
\citep{NVSS} at 1.4 GHz.
The power pattern is shown in gray-scale with a scale bar at the
top and contours are plotted at the $\pm 0.02, \pm 0.04, \dots \pm 0.16$ levels
of fractional Stokes V (from the center outwards).  Negative contours are dashed.
}
\label{vlaBeam}
\end{figure}

Many galactic objects have significant circular polarization
in their emission. 
Without correction for the beam squint, the (E)VLA has strong off-axis
instrumental circular polarization (Figure~\ref{vlaBeam})
which masks any
intrinsic circular polarization.
Correction for the beam squint will result in a significant increase in the sensitivity
of measurements of circular polarization.

(E)VLA observations are performed using dual orthogonal circular polarization
(henceforth RCP and LCP).
It has been known for quite some time that the antenna gain patterns formed
on the sky for these two polarizations are not concentric, having a relative
offset of $0.06 \pm 0.005$ of the antenna power FWHM; in reasonable 
agreement with a calculated value of 0.053 \citep{Napier1977}.  The NVSS survey
included careful measurements of the primary beam response of the VLA antennas
at a frequency of 1.4~GHz.  The beam squint was found to be $1.71' \pm 0.02'$ \citep{NVSS}
corresponding to $ 0.055\ \times$~FWHM (Figure~\ref{vlaBeam}).
Thus, for perfectly pointed observations, the gain in RCP and LCP
towards a source located away from the pointing position will be modulated by up to
$\sim 7$\% due to the Earth's rotation in approximately opposite sense below or
above the nominal gain depending on the location of the source.  The effect will be
approximately equal in magnitude for the RCP and the LCP but
of opposite sign and thus will largely cancel when estimating Stokes I;
as long as both polarizations have been observed and preserved in the
editing process.  However, amplitude self-calibration might magnify the effect if corrections
are determined separately for both polarizations as they will tend to counter the modulation of
the strongest source affected by the beam squint.  This can be prevented by preaveraging the
RCP and LCP data prior to determining common amplitude correction factors.

In this paper, we discuss the beam squint in some detail, present
an algorithm that corrects its effects and
apply it to observations made with the VLA.  We discuss as well the possible
extension of this technique to a wider class of time and/or direction-dependent corrections.
All data manipulation discussed in this report use the Obit package \citep{Cotton2008},
(http://www.cv.nrao.edu/$\sim$bcotton/Obit.html) .

\section {The (E)VLA beam squint}

Extensive studies of the polarization properties of ``offset parabolas'' have been performed since the early 1970s with detailed calculations describing the squint and cross-polarization properties of prime-focus offset parabolas \citep{Chu1973}.
Subsequent, more refined studies determined the polarization properties of a single-reflector parabola
illuminated by a circularly-polarized feed located at the prime focus and whose axis is tilted by an angle
$\theta_0$ with respect to the axis of the parabola.  The resulting antenna primary
beam is directed off-axis
orthogonally to the plane that contains the axis of the parabola as well as that of the feed.  The
beam squint is given by \citep{Duan1991}:

\begin{equation}
\label{eqn1}
{\theta_S}\ = \ sin^{-1} {\biggl(\frac {sin \theta_0 \lambda }{ 4 \pi F }\biggr)}
\end{equation}
where $\theta_0$ is the angle between the feed axis and the axis of the parabola, $\lambda$ is the wavelength of
the observations and $F$ is the focal length of the parabola, ($F = 9.0$~m at the (E)VLA) and
$\mp \  \theta_S$ is the resulting beam squint of the (LCP, RCP) beam.

The dual reflector offset Cassegrain antenna was solved by defining an equivalent prime-focus
parabola that is equivalent in performance to the wanted dual reflector Cassegrain antenna as it has
the same co-polarized and cross-polarized radiation patterns \citep{Rusch1990}.  It can be shown that
the beam squint of such an antenna configuration is given by

\begin{equation}
\label{eqn2}
{\theta_S}\ = \ sin^{-1} \biggl( {\frac {sin ( \theta_\beta - \alpha)  \lambda }{ 4 \pi F_{\rm eq} }\biggr)}
\end{equation}
where $\alpha$ is the angle between the axis of the equivalent parabola and that of the feed and
$\theta_\beta$ is the angle between the feed and the vertex of the off-axis, shaped subreflector and
$F_{\rm eq}$ is the equivalent focal length of the dual reflector.  At the (E)VLA antennas, the
corresponding values are $\alpha = 59.8^\circ$, $\theta_\beta = 1^\circ$ and $F_{\rm eq} = 59.1$~m
\citep{Napier1994}.  This reference discusses the derivation in some detail and shows that the beam
squint is
thus described by an inverse-sine function.  However, the small-angle approximation is appropriate and
thus the squint angle at the (E)VLA antennas scales quasi-linearly with wavelength.  This is not necessarily
the case with the FWHM of the primary beam response as this depends on the specific illumination due
to the feed horns at each band.  The description of the primary beam response as corresponding to
uniform illumination of a 24.5~m circular aperture (appropriate at the VLA at 1.4~GHz given the 25~m
antenna diameter and the taper applied at the edge of the primary) leads to the 5.5\% ratio discussed
above.  We have found that this ratio holds as well for the new (E)VLA 5~GHz feed system (Uson \&
Cotton, in preparation) but do not
know at this time what the corresponding ratio at other frequencies of the (E)VLA will be.  We intend to
perform the relevant measurements in due course.

\section {Beam squint correction technique}
The beam squint can be characterized as a coupled offset of the pointing of the
two orthogonally polarized beams.
The antenna will be nominally pointed using the midpoint of these two
beams giving an (opposite) effective pointing error in each polarization.
In the following, the beam squint is treated as a polarization-dependent pointing error.

The beams are offset orthogonally to the line connecting the feed and the
antenna center.
Due to the alt-az mounts of the (E)VLA antennas, this offset will rotate
on the sky with parallactic angle. 
Notice that by construction of the station pads, all (E)VLA antennas have the
same parallactic angle at any given time even in the extended configurations.

The effect of beam squint on a given source depends on its location in
the field of view so, in general,
no operation on the uv data can remove the effects of beam
squint over the whole field, although an approximation can be attempted through
construction of a suitable unitary operator (e.g.~\citep{Bhatnagar2008}).  However,
such a procedure encounters problems on fields with significant emission from sources
located near the nulls of the primary beam.
The effects of beam squint can be included whenever the instrumental
response to a model (e.g., a set of CLEAN components) is computed.
In the algorithm described here, this is done using a discrete
Fourier transform where the instrumental response
for each visibility is computed for each component and then summed
over components. 
The flux density of each component is modified by the ratio of the
antenna gain in the nominal pointing direction to the actual, beam
squint dependent, gain.
This is done independently for each polarization.
As the beam squint rotates with parallactic angle, this flux density
correction is time as well as direction dependent.

There are two instances where this interferometer response model is used.
The first is in a visibility-based
(``Schwab-Cotton'', \cite{Schwab1983}, \cite{Cotton1989}) CLEAN in which a set of 
components is initially located by a major cycle of the 
``Clark'', \cite{Clark1980} CLEAN followed by an accurate model
calculation allowing the effects of those components to be removed
from the visibility data and a new residual image derived from the
residual visibility data. 
Thus, errors in the initial image due to inadequacies in the initial
response model are corrected as deconvolution progresses.

The other use of the instrumental response model is in the
self-calibration step in which the visibilities are divided by the corresponding
instrumental response to the sky model to produce a dataset equivalent
to observations of a point source with a perfect instrument.
Using an accurate beam-squint model allows the removal of the effects of beam
squint from the resultant gain solution.

If an image in Stokes~V is desired, then Stokes~I must be imaged first 
and the Stokes~I model subtracted from the data with beam squint
corrections applied. 
This should remove the instrumental circular polarization from the data.
Subsequently, the Stokes~V image can be derived and deconvolved
without further beam squint corrections.  
Instrumental leakage terms (the so-called D-terms) have been ignored
in the procedure discussed in this paper, but could be incorporated
if necessary as long as such leakage terms were determined with sufficient precision.

\subsection{CLEAN model for beam squint corrections}
In the normal model calculation, a list of CLEAN components is
kept and, for each visibility measurement, the instrumental response to
each component is calculated with the total instrumental response obtained by
summing over all components.
To apply the beam squint correction, the time- and polarization-dependent
antenna gain corrections are used to adjust the component
flux density for the effects of beam squint on the antenna power
pattern.
To do this, the antenna voltage gain relative to the nominal pointing
is determined for the RCP and LCP beams in the direction of each CLEAN
component and the interferometric (power) gain correction is obtained as the product
of the antenna voltage gain corrections. 
These gain corrections are updated whenever the observing parallactic angle
changes by one degree.  Of course, this interval could be shortened if necessary.

A mixed array such as the VLA+EVLA adds another complication as the
feed placement is different for the two types of antennas \citep{CottonUson2007a}.
In this case, the right- and left-hand voltage gain ratios are computed
separately for VLA and EVLA antennas and the appropriate voltage gains
are applied to the component flux density for each baseline.

The implementation described above assumes that all antennas observe the sky at the same
parallactic angle.  This is indeed the case at the (E)VLA as care was taken during
the construction of the VLA to build the antenna mounts to a common horizon.
For arrays such as the VLBA, parallactic angle will differ from
antenna to antenna and an implementation for such an array
would need to maintain lists of voltage gain corrections for each antenna.
This is straightforward and simply adds to the procedure some bookkeeping
as well as computational overhead.

If large numbers of CLEAN components are involved, this technique is
expensive in computer cycles and memory.
A simple expedient procedure is to ``compress'' the list by summing
the flux density of all components derived from a given image grid
cell and using at most one component per cell.
An additional benefit of summing the components in each cell is the
reduction of numerical round-off error in calculating the model
response. 

\subsection{Beam squint gain corrections}
 This beam squint correcting algorithm has been implemented 
in the Obit package as task ``Squint.''
The pointing offset in antenna coordinates due to the beam squint is
orthogonal to the line from the feed to the antenna center with its value
given by equation 2 using the values given by \cite{Napier1994}

Thus, the offset for each polarization at the (E)VLA is given by
\begin{equation}
\label{eqn3}
{\rm squint}\ = \ \mp 237.56 \times \lambda (\frac{arcsecond}{meter})
\end{equation}
where  $\lambda$ is the wavelength (m).
The value of squint is negative for LCP and positive for RCP.
The offsets in RA and Dec are then:
\begin{equation}
\label{eqn4}
\begin{split}
{\rm dx} &= {\rm squint} * (-sin({\rm feedAngle} - \chi))\\
{\rm dy} &= {\rm squint} * (cos({\rm feedAngle} - \chi))\\
\end{split}
\end{equation}
where feedAngle is the orientation of the feed on the feed circle,
$\chi$ is the parallactic angle and the
sign of squint is appropriate for the polarization \citep{CottonUson2007a}.

For a given CLEAN component, the distance from the pointing position
corrected for beam squint is given by:
\begin{equation}
\label{eqn5}
\begin{split}
d({\reffont x, y})\ = cos^{-1}\{ &sin(\delta_{off}) sin(\delta_{Pnt})\ +  \\
&cos(\delta_{off}) cos(\delta_{Pnt}) cos (\alpha_{off}-\alpha_{Pnt})\}\\
\end{split}
\end{equation}
where $\alpha_{Pnt}$ and $\delta_{Pnt}$ are the RA and Dec of the
antenna pointing, 
$\alpha_{off}\ =\ \alpha_{CC} + {\reffont x}$,
$\delta_{off}\ =\ \delta_{CC} + {\reffont y}$,
and $\alpha_{CC}$  and  $\delta_{CC}$ are the RA and Dec of the
component.

For frequencies above 1 GHz, the antenna pattern is approximated above
the 5\% level by a Jinc function\footnote{The arbitrary cutoff at the 5\%
response is due to our lack of precise knowledge of the (E)VLA primary beam
response at low levels; we are investigating ways to extend the squint correction
to lower levels.} which is in turn approximated using {\reffont \S9.4.4 in} \cite{Abramowitz1964}: 
\begin{equation}
\label{eqn6}
P({\reffont x, y})\ =\ 4.0\, \bigl( 0.5 + \sum_{n=1}^6 c_n u^n \bigr)^2
\end{equation}
where 
$$u\ =\ \bigl\{1.496 \times 10^{-9} \times \bigg(\frac{25.0}{d_{ant}}\bigg)\times d({\reffont x, y}) \times \nu\bigr\}^2  $$
and $c_1$=-0.56249985, $c_2$=0.21093573, $c_3$=-0.03954289, $c_4$=0.00443319,
$c_5$=-0.00031761, $c_6$=0.00001109 and $d_{ant}$ is the effective antenna diameter in
meters with $d_{ant} \sim 24.5$~m at a frequency of 1.4~GHz \citep{NVSS}.
The voltage gain for each antenna and polarization is then:
\begin{equation}
\label{eqn7}
{\rm voltage\_gain}\ =\ \sqrt{\reffont P(x, y)} \ .
\end{equation}

As described above, the correction assumes a single
azimuthally-symmetric antenna pattern describable by a 1-D function. 
A single 2-D antenna pattern could be readily incorporated as a lookup table.
Antenna-specific beam patterns require maintaining separate lists of
component gains for each antenna.
In the case of a large number of components, this method can be time
consuming.  Therefore, in the implementation described here, a threshold
is defined that specifies the cutoff between the levels that are processed
using the high accuracy squint correction model and those that are not.
If the purpose of the application of this technique is
solely to improve the dynamic range in Stokes~I by reducing the
artifacts arising from the brightest emission in the field, it is not
necessary to compute a highly accurate model for components
corresponding to regions of weaker emission. 
By restricting the accurate model calculation to the portion of the
model representing the brightest emission and allowing faster, but
less accurate calculations for the weaker emission, the expense of this
algorithm can be drastically reduced.  Because extended, low-level emission
can produce a coherent error pattern it is important to keep this threshold
rather low.  In addition, 
if Stokes V imaging is desired, the beam squint correction must be
applied to the entire Stokes I model.
During CLEANing, whenever the peak in the initial residual image at
the start of a major cycle exceeds the threshold value, the high accuracy model
is required for all components.
When the peak residual drops below this threshold, a lower accuracy model
calculation is allowed if it is deemed to be faster.
For self-calibration or other cases where the total CLEAN model is
known, the components corresponding to a given image cell are combined
and those whose flux densities exceed the threshold are processed with the high
accuracy model, whereas those below the threshold may be processed with the
faster method.
In all cases, components corresponding to a given cell are combined
before the model calculation.  
A nonzero threshold for using the more accurate calculation might limit
the dynamic range because of its effect on the amplitude
self-calibration; a lower threshold leads to higher accuracy,
although at the expense of execution time. 

\subsection{Correction of individual antenna patterns}
Detailed differences among antennas will cause differences in their
radiation patterns.
For extremely high dynamic range, wide-field observations corrections
for these differences among antennas will be necessary.
This case is similar to the beam squint problem in that time-variable and
directional gain corrections are needed.
A straightforward modification of the technique described in this paper would
allow correcting for detailed antenna patterns, so that errors due to cross-polarization
leakage could be corrected in the manner described here.

\subsection{ Correction of antenna pointing errors}
As mentioned in the introduction, in the absence of pointing errors,
the effects of beam squint on RR and LL correlations will be
approximately equal but of opposite sign.
Thus, in this case, imaging of Stokes I (0.5 *(RR+LL)) will be
largely, but not completely, insensitive to beam squint.
However, in the presence of pointing errors, the magnitude of the beam
squint on RR and LL will be different and will not be eliminated 
by imaging Stokes I, or by the technique described here without
additional pointing corrections.

\begin{figure*}
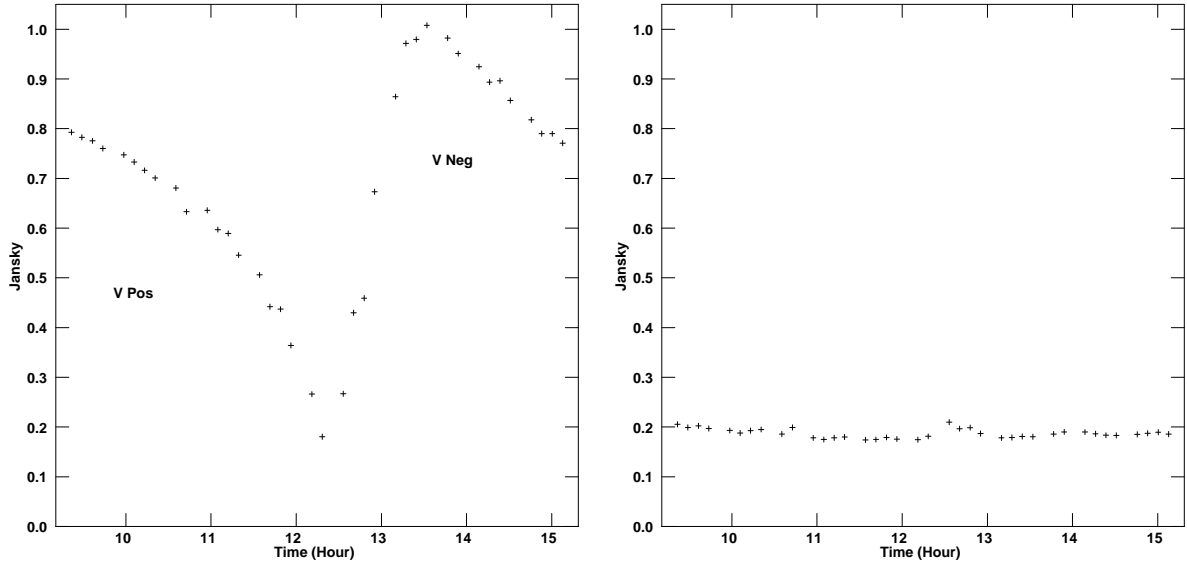

\centering
\centerline{
   \includegraphics[height=3.25in]{9509fg2a.eps}
   \includegraphics[height=3.25in]{9509fg2b.eps}
}

\caption{ 
Time averaged 1.4 GHz Stokes~V for observations with 3C84 located at
the half-power point westward of the pointing center.
The left panel shows uncorrected data, the minimum is located where the Stokes~V
changes sign; positive and negative regions of Stokes~V are indicated.
The right panel shows the corresponding data after subtraction of the beam-squint corrected
Stokes~I model. 
The amplitude bias of the individual measurements is $\sim 200$~mJy.
}
\label{VPlots}
\end{figure*}

At this writing, global, antenna-based pointing errors (i.e., collimation errors) can be
readily incorporated as our algorithm treats the beam-squint as a coupled (R,L)
pointing error.
If post-observation corrections of time-variable antenna pointing errors could be
determined, a straightforward modification of this technique could be applied.
Effective component gains for each antenna and polarization could be kept,
including the antenna pointing errors as well as the beam squint.
A time and antenna dependent addition to the beam squint position
offsets would be used in computing the effective component gains.  This would
simply require some bookkeeping and involve extra, straightforward computations.

\section {Verification using {\reffont actual} data}

\subsection{Example 1: Off-axis observations of 3C84}

The beam squint algorithm was tested on VLA B configuration
observations of 3C84 in which the source was alternatively located on-axis,
at the 1/2~power of the antenna beam (westward) and at the
1/2~power and 1/3~power of the antenna beam (southward).
Observations were made in both the 1.4 and 5 GHz bands in spectral line mode,
recording 15 $\times$ 390 kHz channels.
The observations were made in July and August 2006.
The data were amplitude, phase and bandpass calibrated in the normal
fashion using the on-axis pointings with 2-point interpolation of the gain solutions
(amplitude, phase) to the other pointings

The data were processed twice, once without and once with beam squint
correction applied but otherwise processing was the same.
All imaging used the ``autocenter'' technique \citep{CottonUson2007b}, to
improve the dynamic range.  A single iteration of phase-only
self-calibration with a solution interval of 30~seconds
was followed by a single iteration of amplitude and phase
self-calibration with a 2-minute solution interval. 
Self-calibration solutions were determined for an average of right- and
left-hand circular polarization as without beam squint corrections,
the independent right- and left-hand amplitude self-calibration would
``correct'' the data for 3C84 at the expense of the rest of the field.
{\reffont In the following paragraphs, we discuss the observations of 3C84 at a
frequency of 1.4~GHz with the source observed at the (westerly) half-power
location of the antenna primary beam}.

A convenient way to see the effects of beam squint and to test the
efficacy of the correction is to examine a plot of the time-averaged
Stokes~V amplitudes averaged as well over baselines and spectral channels
(Figure \ref{VPlots}, {\reffont left panel}).
The instrumental polarization is fixed to the antenna and will rotate on the sky
with parallactic angle causing a time-variable Stokes~V signal in the
averaged data.
Since the source is not located at the phase center,
this averaging must be done over amplitudes and therefore will suffer
from a significant amplitude bias.
Also, since the figure shows only the amplitude of the averaged visibilities,
both positive and negative Stokes~V appear with positive sign with a dip in the curve where the 
instrumental response changes the sign of Stokes~V.

Each data set was imaged using
these corrections to derive a Stokes~I model which was then
subtracted from the data after performing the squint correction.
A time-averaged Stokes~V plot (Figure \ref{VPlots}, {\reffont right panel}) will
show the extent to which the procedure has worked.
The 200~mJy level in  Figure \ref{VPlots} corresponds to the amplitude
bias of the individual time and spectral samples.
This figure shows that the beam squint has been corrected to the level expected from the amplitude
bias and within the scatter expected from the noise in the individual measurements. 
Similar results were obtained for the 5~GHz data.

\begin{figure*}
   \centering
\centerline{
   \includegraphics[height=3.25in]{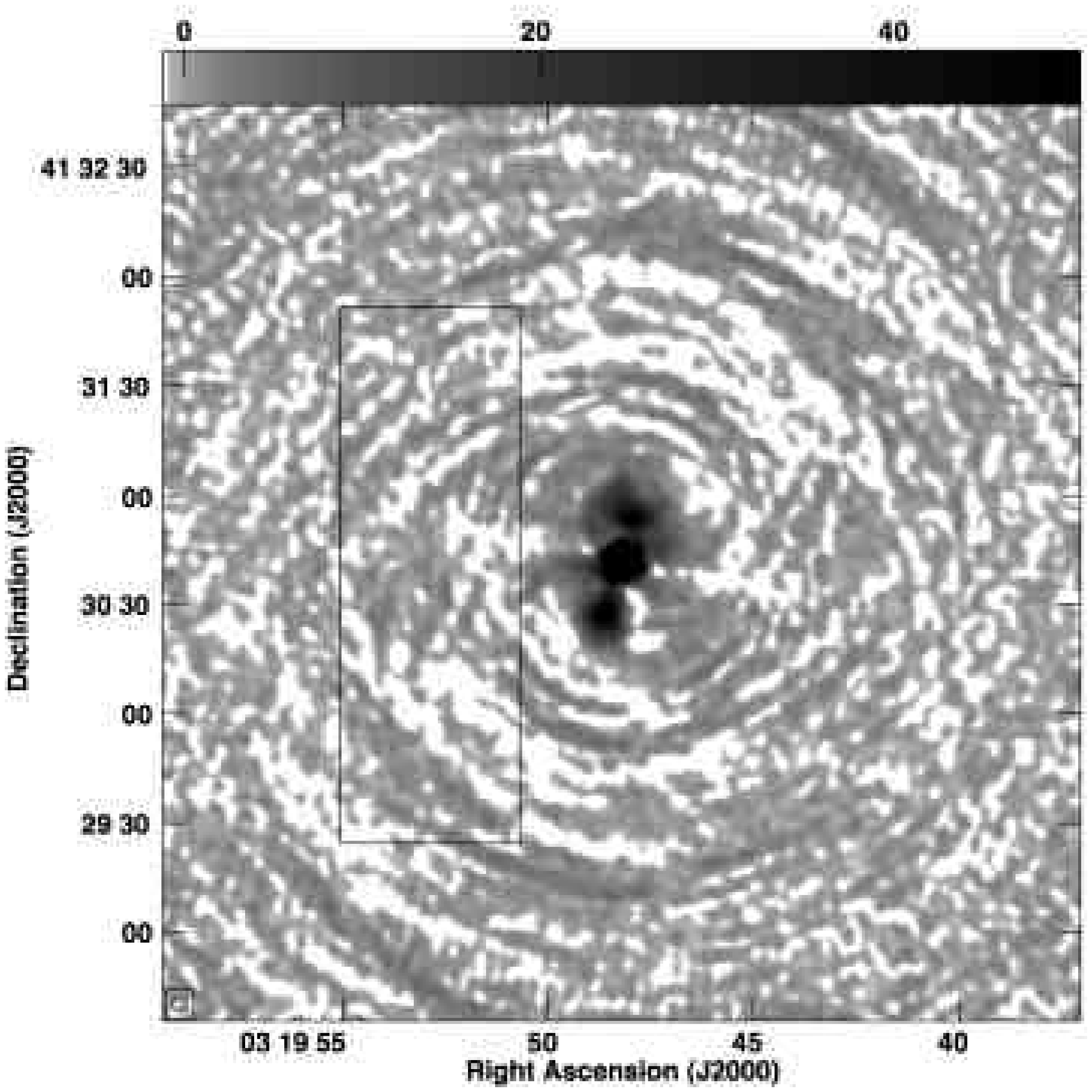}
   \includegraphics[height=3.25in]{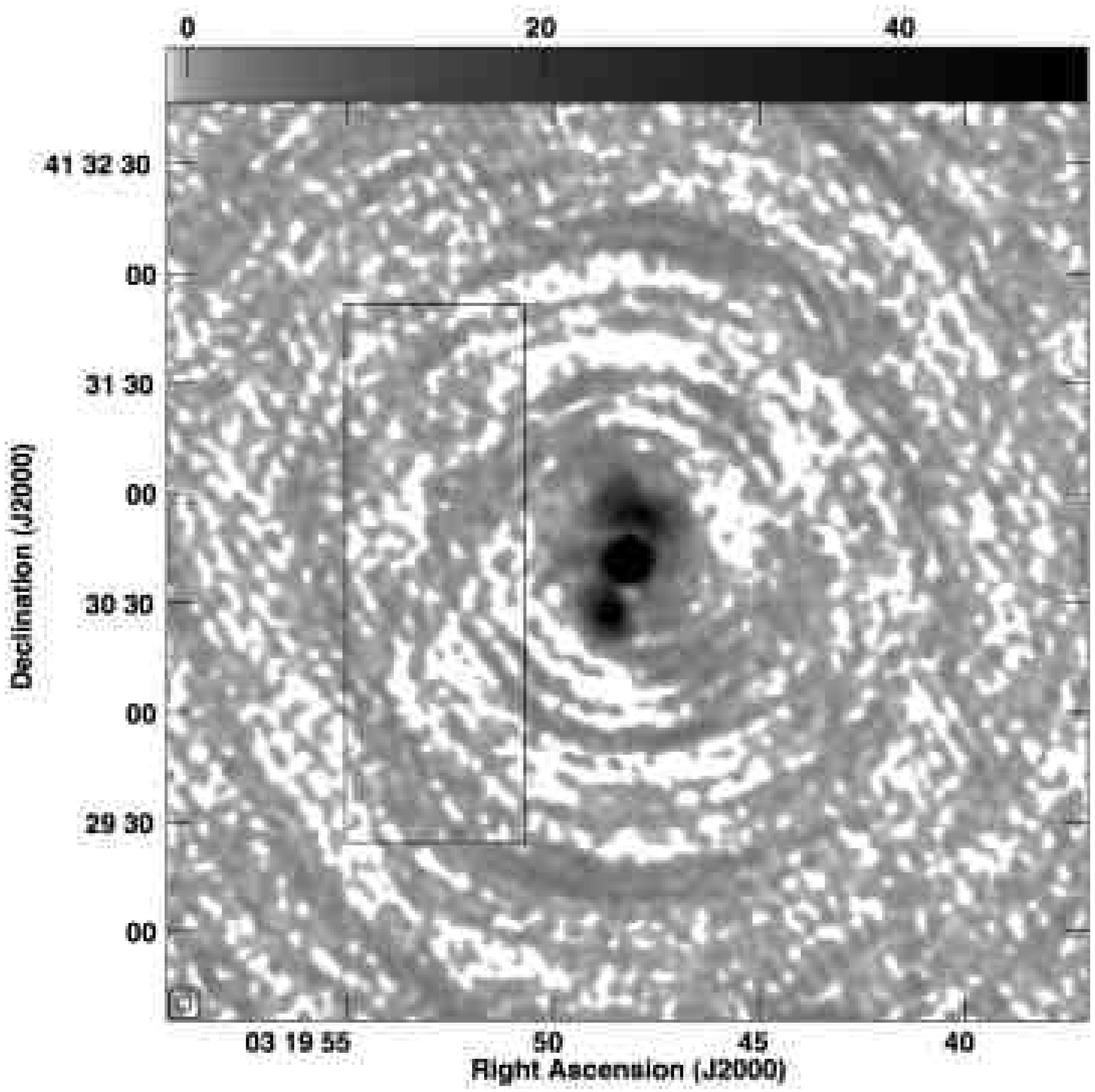}
}
\caption{ 
{\bf Left:}  Example 1. Stokes I in reverse gray-scale of the region around 3C84
imaged with no beam squint correction. 
The image is displayed with a square root transfer function and the
scale is given by the wedge (in mJy/beam) at the top.
The box gives the region over which the RMS was determined.
\hfill\break
{\bf Right:} As left but with beam squint corrections.
}
\label{IImage}
\end{figure*}
\begin{figure*}
   \centering
\centerline{
   \includegraphics[height=3.5in]{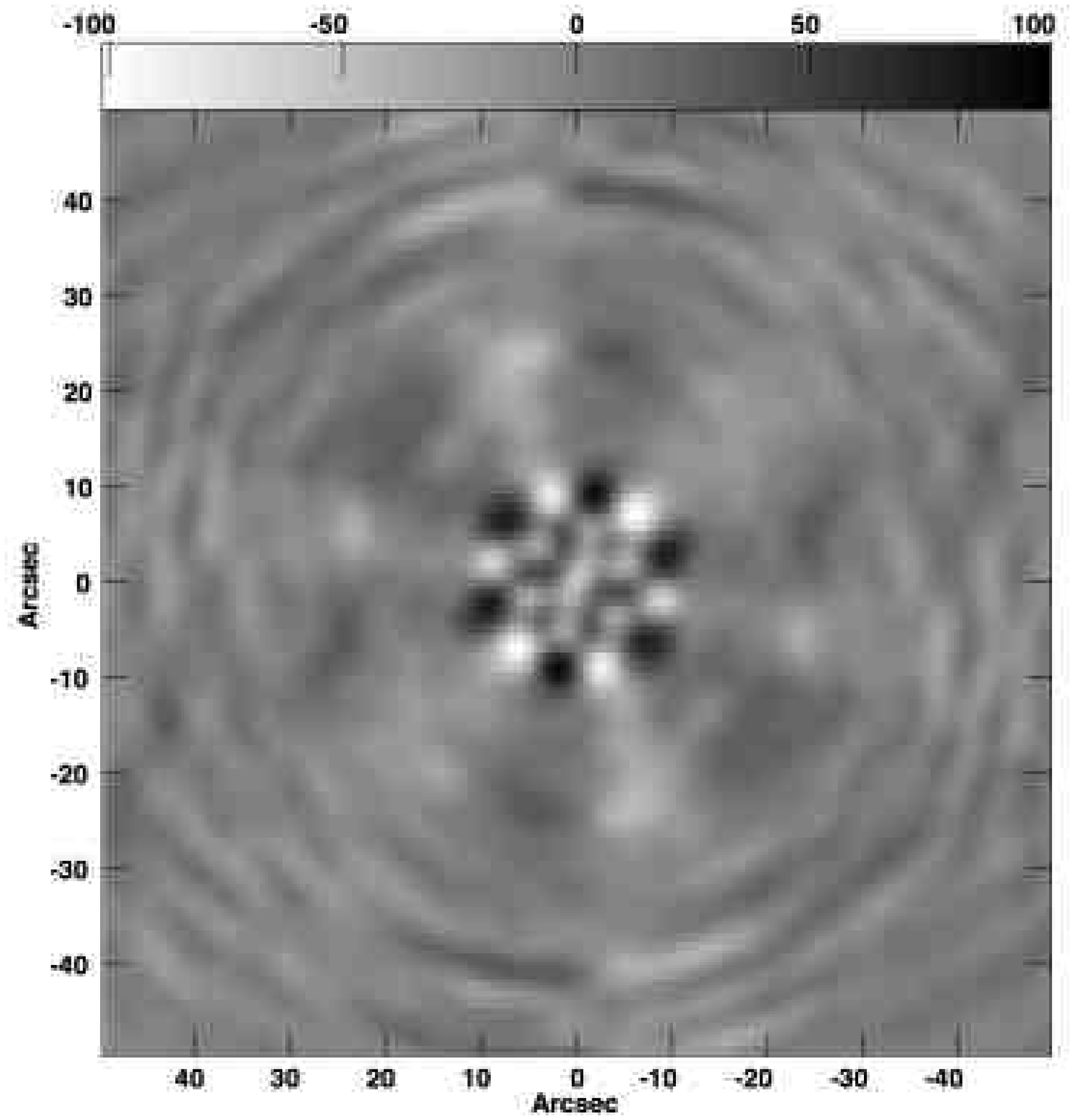}
   \includegraphics[height=3.5in]{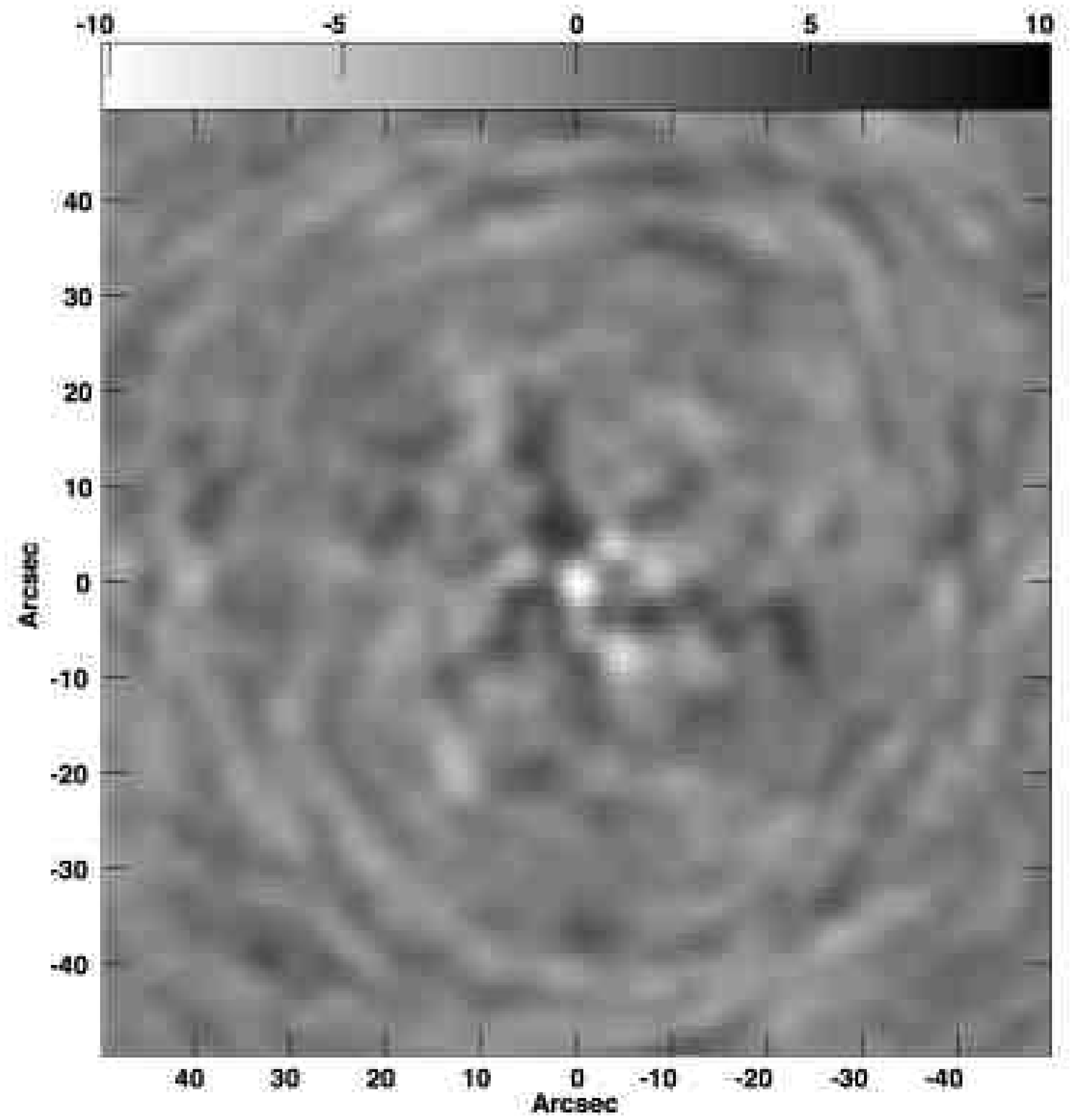}
}
\caption{ 
{\bf Left:} Example 1. Stokes V in reverse gray-scale of the region around 3C84
imaged with no beam squint correction shown  with a linear display of
-100 to 100 mJy/beam.
The scale is given by the wedge at the top.
\hfill\break
{\bf Right:} As left but with beam squint corrections.  {\reffont Notice the finer scale
displaying the range of values -10 to 10~mJy/beam}. 
}
\label{VImage}
\end{figure*}

The Stokes~I images are shown in Figure \ref{IImage} with a square-root
stretch of the lower pixel values in the image (peak 12.2 Jy/beam).
The image made without beam-squint correction shows more residual
artifacts than the image with the correction applied.
To compare the two images quantitatively, the RMS noise was computed in
identical boxes which are shown in Figure \ref{IImage}.
The RMS in the image without correction was 1.47~mJy/beam whereas it
was 1.16~mJy/beam in the image made with the
squint correction, a reduction of the level of
artifacts by 21\%.

Figure \ref{VImage} compares the Stokes~V results.
Since the data included observations over a range of parallactic
angles, much of the instantaneous instrumental circular polarization is
washed out.  {\reffont Nevertheless, the image made without squint correction
shows residual errors that are one order of magnitude larger than those
in the squint-corrected image which is displayed using only the inner 10\% 
of the pixel range of the uncorrected image.}
In the uncorrected image, the maximum and minimum values are well off
the source but the maximum value in the corrected image is on-source.
However, this analysis becomes confused as the core in this source is known to
show circular polarization above the level shown in Figure \ref{VImage}.
Since the data were calibrated using 3C84 and assuming no circular
polarization, the polarization calibration is incorrect.
Unfortunately no other calibrators were included in the observations.
The off-source noise levels computed using the same region as in the Stokes~I
images are 7.9~mJy/beam and 0.94~mJy/beam for the un-corrected and corrected
images (respectively).

In this test, the beam squint correction leads to a significant
improvement to the quality of the Stokes~I image and to a very large
improvement in the Stokes~V image.

\subsection{Example 2: Imaging of full-track observations}
\begin{figure*}
   \centering
\centerline{
   \includegraphics[height=3.25in]{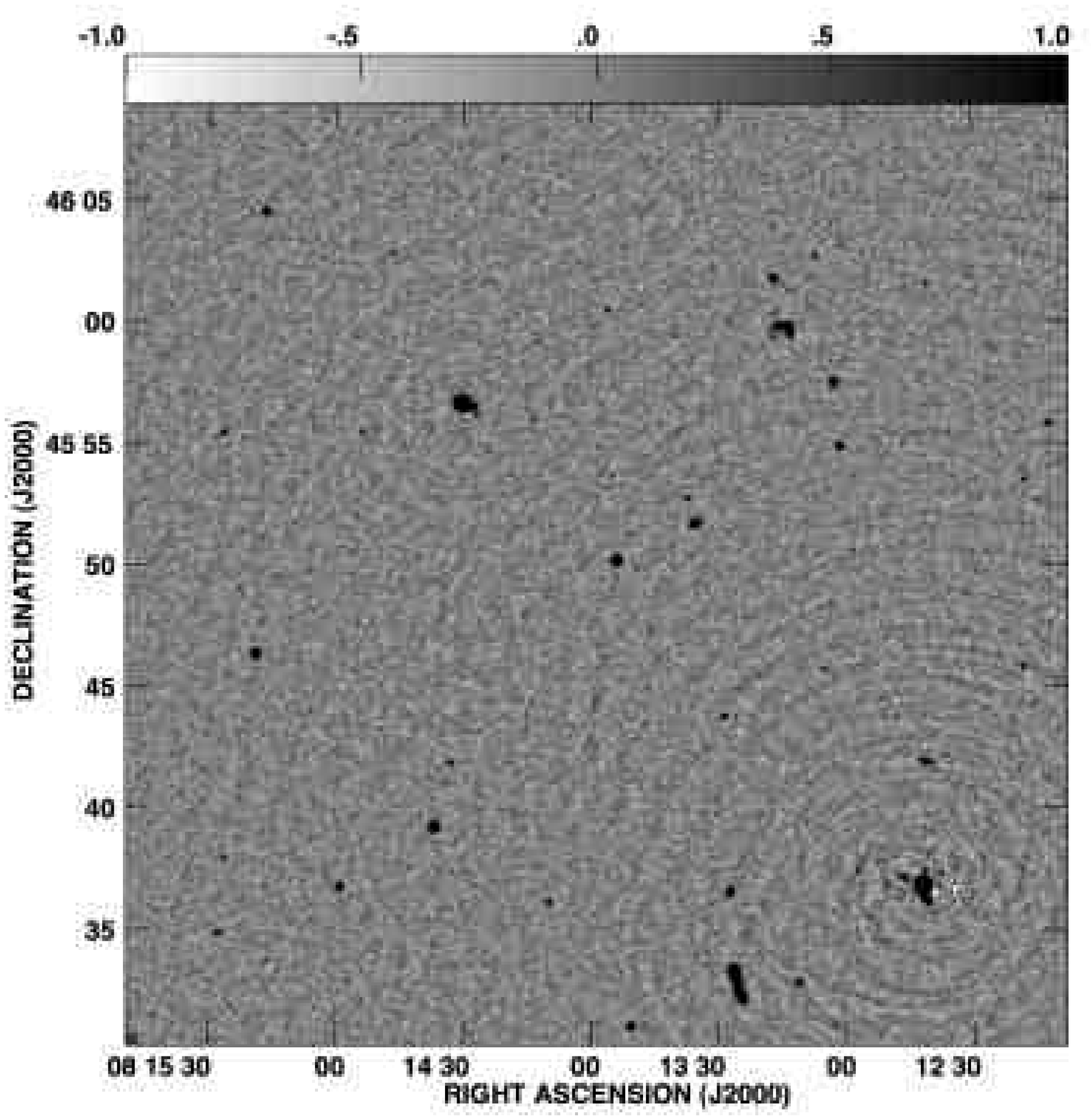}
   \includegraphics[height=3.25in]{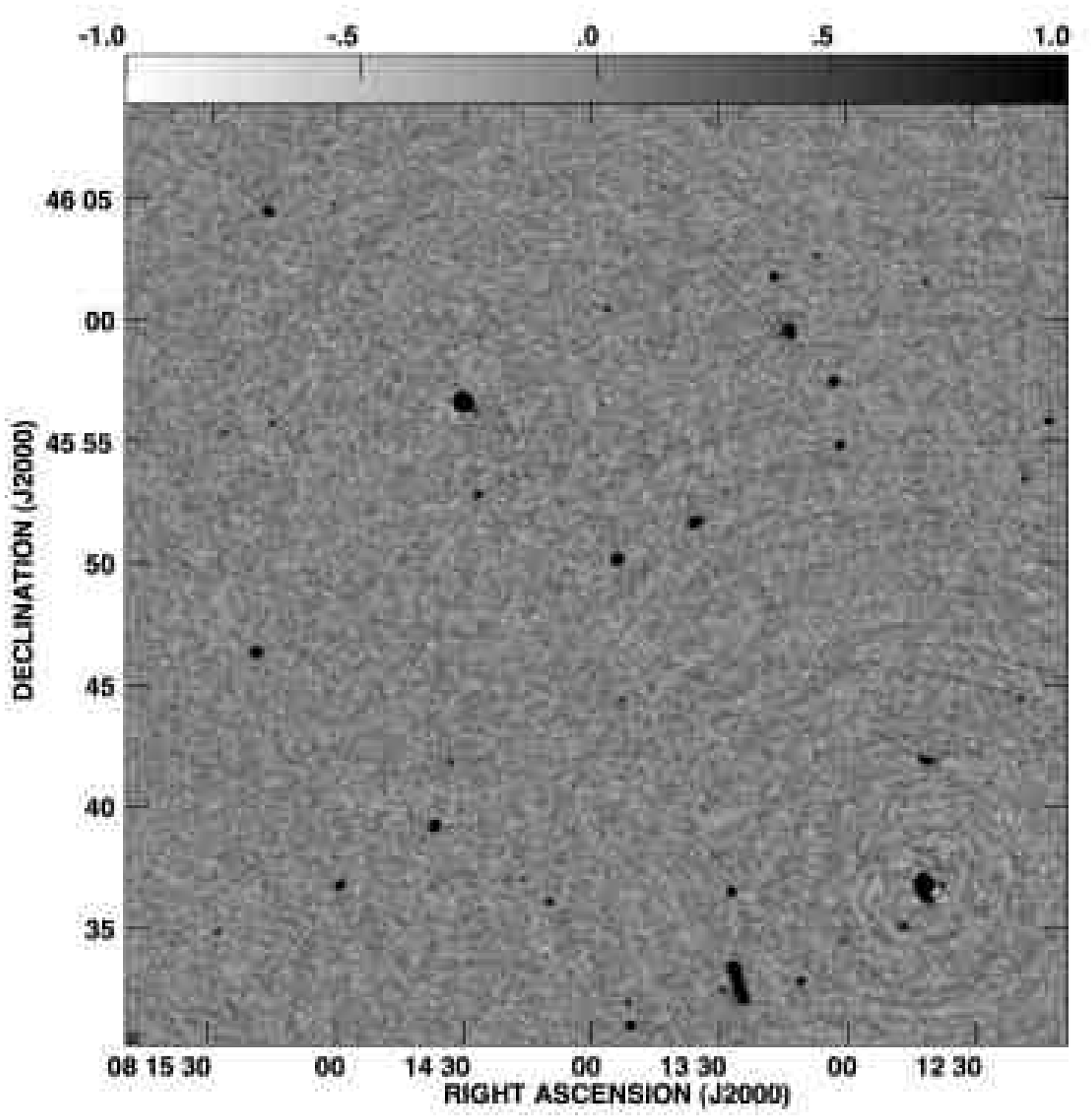}
}
\caption{ 
{\bf Left:}  Example 2. Stokes I in reverse gray-scale of the field of IC~2233
imaged with no beam squint correction.  Notice the deconvolution errors due to the two
``4\ C'' radio sources.
The image is displayed with the scale given by the wedge (mJy/beam) at the top.  The rms noise is
$\sim 150\ \mu$Jy/beam.
\hfill\break
{\bf Right:} As left but with beam squint corrections.  The noise is reduced to $\sim 110\ \mu$Jy/beam.
Some artifacts remain that are centered on the ``4~C'' source on the lower-right side.
}
\label{IImage2}
\end{figure*}
\begin{figure*}
   \centering
\centerline{
   \includegraphics[height=3.25in]{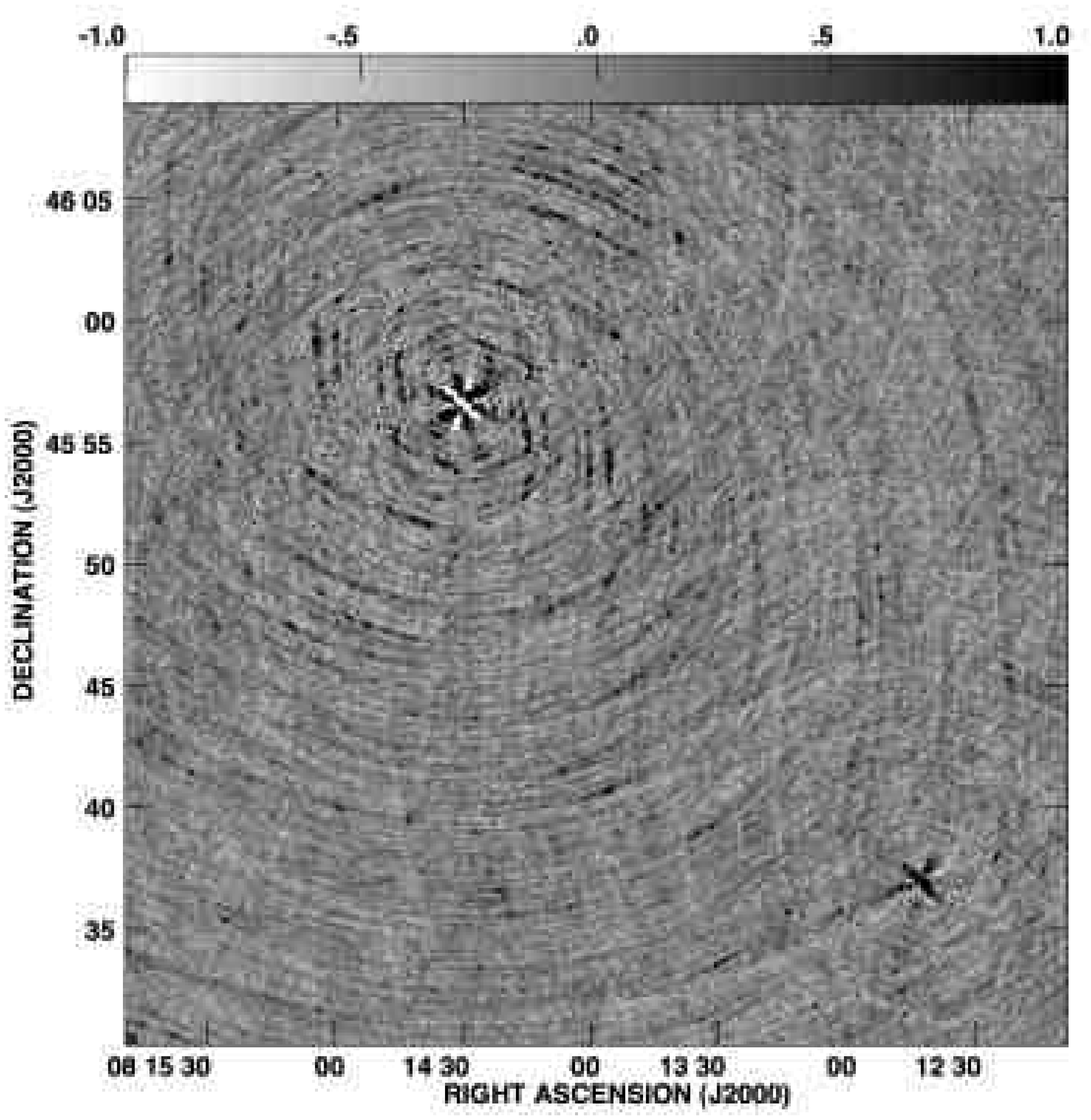}
   \includegraphics[height=3.25in]{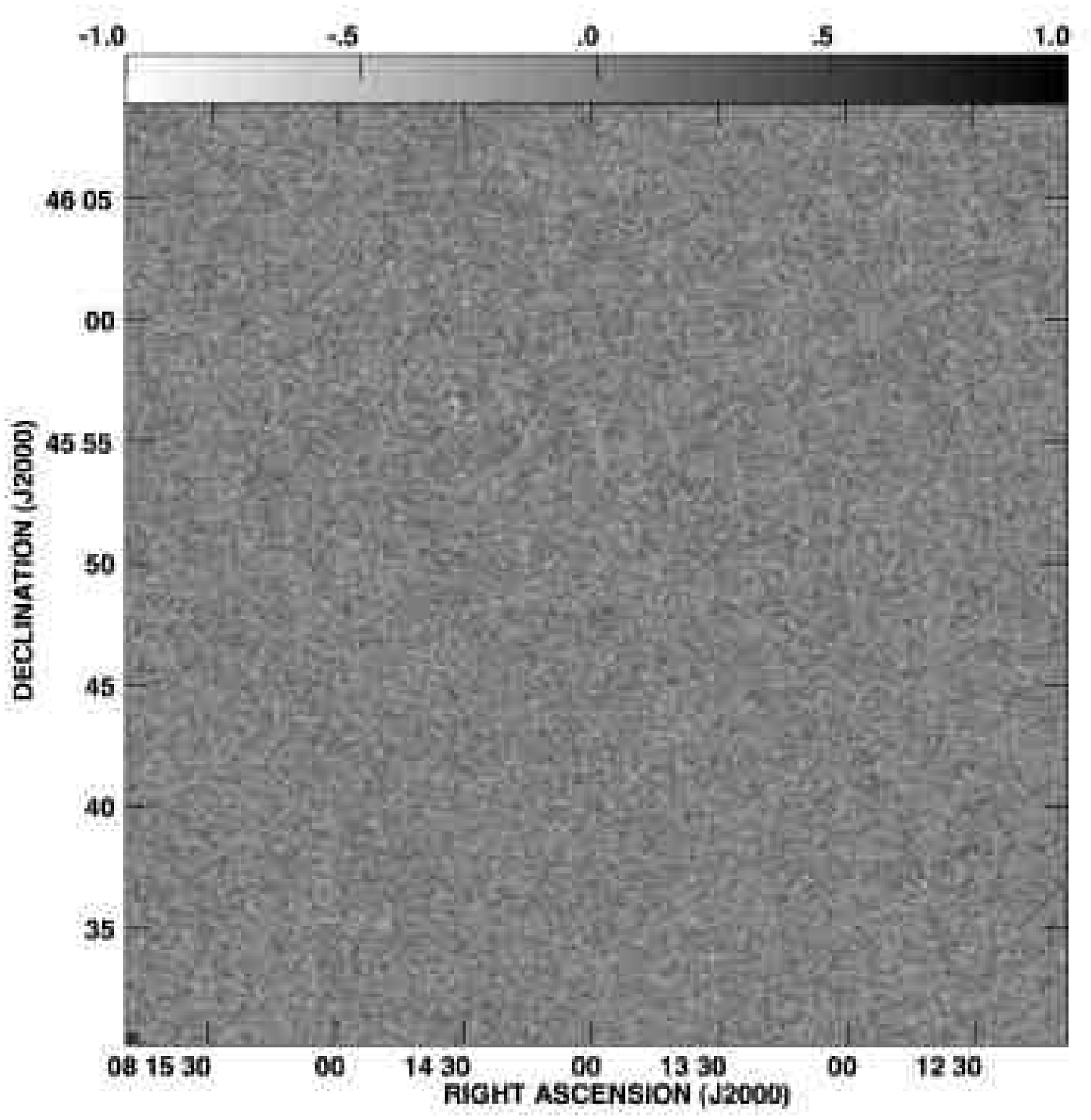}
}
\caption{ 
{\bf Left:} Example 2. Stokes V in gray-scale of the field of IC~2233
imaged with no beam squint correction.  Notice the artifacts due to the variable Stokes~V
``signal'' at the position of the two ``4\ C'' sources induced by the Earth's rotation of the
(squinted) primary beam response with respect to the sky.  This error pattern determines the
RMS noise value of $190\ \mu$Jy/beam.
The scale is given by the wedge (mJy/beam) at the top.
\hfill\break
{\bf Right:} As {\reffont left} but with beam squint corrections.  The rms noise is $\sim 105\ \mu$Jy/beam.
}
\label{VImage2}
\end{figure*}

Our second example is a deep, full-track (two passes) {\reffont 1.4~GHz H\,{\sc i}}
observation of the field of the
superthin galaxy IC2233 which contains significant background continuum emission totaling
$\sim 1.1$~Jy (observed, i.e. before correction for the attenuation of the primary beam) which
is dominated by two bright ``4C'' sources (4C+46.17 at
J081430.4+455639.4 
(henceforth the ``upper-left'' source) and 4C+45.15 at 
J081242.0+453651.3 (hereafter the ``lower-right'' source).
The sources are observed with
flux densities of 0.87~Jy and 0.18~Jy, uncorrected for the 80\% and 34\% (respectively)
attenuation of the primary beam.  These sources limit the dynamic range significantly.
The data were obtained with the VLA in C configuration
on 2000 May 28-29 during a period of solar maximum.
A full analysis of the spectral observations
has been published elsewhere \citep{MatthewsUson2008}.  We
discuss here the imaging of the continuum emission with and without the squint correction
as well as the limiting factors and residual errors in the squint-corrected images.

The observations were made in two overlapping frequency bands, {\reffont each with $63 \times
24.4$~kHz channels,} which allowed for 6 line-free
spectral channels from the low-frequency side of the first (lower frequency) band and 14 line-free channels from the high-frequency side of the second (higher frequency) band, after
discarding 6 channels on each end that were contaminated by ghost images from the two strong
``4~C'' sources in the field \citep{Uson2007}.  The remaining line-free channels (6+14) were
processed using conventional methods to make the images shown on the left-hand side of
figures \ref{IImage2} and \ref{VImage2}.  Phase-only self-calibration was
performed on a time scale of 20~seconds with the corrections applied to the data, followed by
amplitude and phase self-calibration performed on 5-minute time intervals after pre-averaging
the RCP and LCP data with these corrections applied to both polarizations.  The Stokes~I image
has an RMS noise level of $150\ \mu$Jy/beam and quite significant
residual sidelobe patterns surround the two ``4~C'' sources.  The variable squint results in a
modulation of the apparent Stokes~V signal as a function of time.  Thus, the uncorrected Stokes~V
image shown in the left-hand panel of figure~\ref{VImage2} shows strong sidelobes that cannot be
deconvolved
as they correspond to a variable ``source'' that does not satisfy the average deconvolution equation
expected from the synthesized beam that corresponds to the full uv-coverage.  The RMS noise is
$190\ \mu$Jy/beam but the noise distribution is clearly non-Gaussian.

We have applied our squint-correcting procedure to these data.  The procedure included
automatic selection of cell-size and imaging grid of ``fly's-eye'' fields and outliers, as well as
``auto-centering'' \citep{CottonUson2007b}, to improve the dynamic range.  The procedure
incorporates phase-only self-calibration (20-second intervals) followed
by amplitude and phase self-calibration (5-minute intervals).  It is worth noting that
the self-calibration operations
used about one half of the number of degrees of freedom used with the conventional analysis
{\reffont because both bands were combined into one 86-channel ``band'' before applying our
algorithm}.  The RMS noise in the resulting images is $110\ \mu$Jy/beam for the Stokes~I image
and $105\ \mu$Jy/beam for the Stokes~V image.

The results are shown in the right-hand panels of figures \ref{IImage2} and~\ref{VImage2}.
Although the Stokes~V image is largely consistent with noise with peak values at
$\pm 4.7 \sigma$, a faint pattern of residual
sidelobes is visible surrounding the position of the upper-left ``4 C'' source. 
Some residual sidelobes to the second-strongest source remain in the Stokes~I image as well.  This lower-right source is located at a point of steep gradient in the primary beam response and is thus
significantly affected by collimation and pointing errors.  We believe that such errors are likely to be
responsible for our failure to correct this image in full.  Of course, any uncorrected systematic effects
will lead to errors in the self-calibration solutions.  Despite the presence of such residual
systematic errors, the Squint algorithm yielded an increase in dynamic range by more than one order
of magnitude in Stokes~V and a decrease of the rms noise in the Stokes~I image of $\sim 27$\%.

\section {Conclusions}
Taking advantage of the EVLA's increased sensitivity at lower
frequencies  (15 GHz and below)
will generally require imaging and deconvolving the
entire primary beam, or at least those portions containing sources.
Beam squint can produce artifacts well above the noise level.
Other interferometers with off-axis feeds, including focal plane
array feeds, will share this problem

This paper describes a technique for correcting radio interferometric
observations for the effects of beam squint.
The technique is a modification of the visibility-based CLEAN in which
the Stokes I image is made in the usual way, but when the image model
is subtracted from the data, corrections are made such that the model
subtracted includes the effects of beam squint.
Thus, any errors incurred in the initial imaging are corrected through
iterative imaging and accurate model subtraction.  Self-calibration is incorporated
into the procedure in two steps with phase-only self-calibration performed
first, subsequently followed by amplitude and phase self-calibration.

We have performed several tests of an implementation of this technique in the Obit task
Squint.  In all cases, careful initial calibration is critical.
Observations of the strong source 3C84 were made with the source at
various locations in the antenna beam pattern resulting in strong
instrumental circular polarization.
Both 1.4 and 5 GHz test data sets were examined and we found that this technique  
corrects the instrumental circular polarization due to the beam squint to
a level that is limited by other effects such as pointing errors and insufficient
knowledge of the primary beam. 

We have presented detailed examples of imaging of several observations.  {\reffont Overall,}
applying the beam squint correction lowered residual artifacts {\reffont in the vicinity of strong
sources} by more than 20\% and led to very substantial improvements to the Stokes~V images.
Our second example demonstrates the behavior of the algorithm on an observation that
requires high dynamic range.  Here, artifacts remain that can be ascribed to collimation and
pointing errors.  It would be straightforward to extend this technique to correct for individual
antenna patterns and known (i.e., determined elsewhere) antenna pointing errors. 

We are investigating the primary beam and pointing parameters of the (E)VLA antennas
at several frequency bands and hope to achieve a more accurate characterization of the squint
correction which should lead to improved correction of the effects discussed here as well as of
the next order of cross-polarization (leakage) corrections.  We hope to report on those results
in due course.

\begin{acknowledgements}
We thank Peter Napier, Jim Ruff and Ken Sowinski for numerous discussions on the actual
characteristics of the (E)VLA antennas, optics and feed systems.  We are indebted to Lynn
Matthews for her excellent bandpass calibration of the observations of IC2233 which required
the computation and application of low-order polynomial corrections as a function of time.  We would
like to acknowledge as well useful discussions with Tim Bastian, Sanjay Bhatnagar, Walter Brisken,
Jim Condon, Tim Cornwell, Rick Fisher, Frazer Owen and Rick Perley.
{\reffont We are grateful to Bob Sault (the referee) for pointing out an error in our original
equation~7 and for suggestions that have allowed us to present a shorter, more concise
description of our examples and conclusions.}
\end{acknowledgements}

\bibliographystyle{aa} \bibliography{9509}

\newcommand{\noopsort}[1]{} \newcommand{\printfirst}[2]{#1}
  \newcommand{\singleletter}[1]{#1} \newcommand{\switchargs}[2]{#2#1}
\begin{thebibliography}{16}
\expandafter\ifx\csname natexlab\endcsname\relax\def\natexlab#1{#1}\fi

\bibitem[{{Abramowitz} \& {Stegun}(1964)}]{Abramowitz1964}
{Abramowitz}, M. \& {Stegun}, I.~A. 1964, {Handbook of Mathematical Functions}
  (Washington, [National Bureau of Standards])

\bibitem[{{Bhatnagar} {et~al.}(2008){Bhatnagar}, {Cornwell}, {Golap}, \&
  {Uson}}]{Bhatnagar2008}
{Bhatnagar}, S., {Cornwell}, T.~J., {Golap}, K., \& {Uson}, J.~M. 2008,
  A\&A, submitted

\bibitem[{{Chu} \& {Turrin}(1973)}]{Chu1973}
{Chu}, T.-S. \& {Turrin}, R.~H. 1973, IEEE Trans. Antennas Propagat., AP-13,
  339

\bibitem[{{Clark}(1980)}]{Clark1980}
{Clark}, B.~G. 1980, A\&A, 89, 377

\bibitem[{{Condon} {et~al.}(1998){Condon}, {Cotton}, {Greisen}, {Yin},
  {Perley}, {Taylor}, \& {Broderick}}]{NVSS}
{Condon}, J.~J., {Cotton}, W.~D., {Greisen}, E.~W., {et~al.} 1998, Astron. J.,
  115, 1693

\bibitem[{{Cotton}(1999)}]{Cotton1989}
{Cotton}, W.~D. 1999, in Astronomical Society of the Pacific Conference Series,
  Vol.~6, ``Synthesis Imaging in Radio Astronomy II'', ed. R.~A. {Perley},
  F.~R. {Schwab}, \& A.~H. {Bridle}, 233--246

\bibitem[{{Cotton}(2008)}]{Cotton2008}
{Cotton}, W.~D. 2008, PASP, 120, 439

\bibitem[{{Cotton} \& {Uson}(2007{\natexlab{a}})}]{CottonUson2007a}
{Cotton},~W.~D.~\&~{Uson},~J.~M.~2007{\natexlab{a}},~EVLA~Memo~113,
  ftp://ftp.cv.nrao.edu/NRAO-staff/bcotton/Obit/Squint.pdf

\bibitem[{{Cotton} \& {Uson}(2007{\natexlab{b}})}]{CottonUson2007b}
---. 2007{\natexlab{b}}, A\&A, submitted

\bibitem[{{Duan} \& {Rahmat-Samii}(1991)}]{Duan1991}
{Duan}, D.-W. \& {Rahmat-Samii}, Y. 1991, IEEE Trans. Antennas Propagat., 39,
  612

\bibitem[{{Matthews} \& {Uson}(2008)}]{MatthewsUson2008}
{Matthews}, L.~D. \& {Uson}, J.~M. 2008, AJ, 135, 291

\bibitem[{{Napier}(1994)}]{Napier1994}
{Napier}, P.~J. 1994, MMA Memo, 115, 1

\bibitem[{{Napier} \& {Gustincic}(1977)}]{Napier1977}
{Napier}, P.~J. \& {Gustincic}, J.~J. 1977, in ``Digest of the IEEE
  International Symposium'', Stanford: IEEE Antennas and Propagation Society,
  452--454

\bibitem[{{Rusch} {et~al.}(1990){Rusch}, {Prata}, {Rahmat-Samii}, \&
  {Shore}}]{Rusch1990}
{Rusch}, W. V.~T., {Prata}, J.~A., {Rahmat-Samii}, Y., \& {Shore}, R.~A. 1990,
  IEEE Trans. Antennas Propagat., 38, 1141

\bibitem[{{Schwab}(1983)}]{Schwab1983}
{Schwab}, F.~R. 1983, in ``Indirect Imaging'', ed. J.~A. {Roberts}, Cambridge
  University Press, Cambridge, England, 333--346

\bibitem[{{Uson}(2007)}]{Uson2007}
{Uson}, J.~M. 2007, VLA Test Memo 237

\end{thebibliography}
\end{document}